\documentclass[preprint]{ptephy_v1}

\preprintnumber{XXXX-XXXX} 
\usepackage{hyperref}
\usepackage{lineno}

\begin{document}

\title{Gamma and neutron separation using emission wavelengths in Eu:LiCaI scintillators}


\author{Takashi Iida}
\affil{Faculty of Pure and Applied Sciences, University of Tsukuba,  Tsukuba, Ibaraki, 305-8571, Japan
\email{tiida@hep.px.tsukuba.ac.jp}}

\author[2,3]{Masao Yoshino}
\author[2,3]{Kei Kamada}
\affil[2]{New Industry Creation Hatchery Center, Tohoku University, Sendai, Miyagi 980-8579, Japan}
\affil[3]{Institute for Material Research, Tohoku University, Miyagi 980-8577, Japan}

\author[4]{Rei Sasaki}
\author[4]{Ryuga Yajima}
\affil[4]{Department of Materials Science, Graduate School of Engineering, Tohoku University, Sendai, Miyagi 980–8579, Japan}


\begin{abstract}%
Scintillators have long been known as radiation detectors and are still used in various applications. Recently, scintillators containing $^6$Li have been developed as neutron detectors and have attracted attention.
$^6$Li absorbs thermal neutrons and emits $\alpha$+$^3$H, which is promising as a neutron detector if it can be separated from background gamma rays. We have been developing Eu:LiI-CaI$_2$-based scintillators (Eu:LiCaI) for this purpose.
In scintillator detectors, waveform information is generally used to distinguish particles such as neutrons and gamma rays. We propose a new particle identification method using emission wavelengths information.
In this study, experiments were conducted using Eu:LiCaI crystals, multi-pixel photon counter optical sensors, and long-wavelength cut filters to verify the proposed method.
The results of irradiating a $^{252}$Cf neutron source and a $^{60}$Co gamma-ray source indicate that there is a particle dependence of the output signal ratio between with and without filters. This indicates that different types of radiation particles have different emission wavelengths.
This is the first demonstration of a wavelength-based particle identification method.

\end{abstract}

\subjectindex{xxxx, xxx}

\maketitle

\section{Introduction}
Scintillator-based radiation detectors are used in various applications because of their ease of use and versatility. 
The development of new scintillators has been actively pursued in past decades \cite{1}.
Recently, scintillators containing $^6$Li have been developed for use in neutron detectors \cite{2} due to the price hike and the shortage of $^3$He.
$^6$Li has a large capture cross-section for thermal neutrons, and the reaction $^6$Li + n $\rightarrow$ $\alpha$ + $^3$H occurs.
The total energy emitted in the reaction is 4.78 MeV, producing an $\alpha$-ray of 2.05 MeV and $^3$H of 2.73 MeV.
If alpha particles and tritium emitted by this reaction can be measured and the background gamma rays can be eliminated, it could be a promising next-generation neutron detector to replace  $^3$He detectors.

In scintillator detectors, the pulse shape discrimination method, which uses the difference in signal waveforms resulting from the difference in dE/dX by particle type, is commonly used to identify the detected particles \cite{3,4,5,6}.
Depending on the type of scintillator, this method typically results in insufficient separation due to small waveform variations.
Therefore, in this study, we investigated a new particle identification method based on data on the wavelength of scintillation light. There are two types of emission mechanisms in scintillators: emission at the host crystal and emission at dopants called activators.
It was suggested approximately 10 years ago that the emission wavelength of Na:CsI crystals may differ depending on the particle \cite{7}, but no detailed study has been reported since then.

In this study, we used a newly developed Eu:LiI-CaI$_2$-based scintillators (Eu:LiCaI) that contains $^6$Li.
We have been developing a scintillator, CaI$_2$, to investigate the nuclear reaction of double beta decay of $^{48}$Ca isotopes.
We have grown CaI$_2$ crystals using the Bridgman–Stockbarger (BS) method
and measurements with a photomultiplier tube have revealed high light yield (approximately 100,000 photons/MeV) and good particle discrimination ability using waveforms \cite{5,8}.
In this study, to take advantage of the neutron capture performance of $^6$Li and the high light yield property of CaI$_2$, we prepared a composite compound of CaI$_2$ and LiI, Eu:LiCaI, and investigated whether neutrons and gamma rays can be discriminated using their emission wavelengths.

\section{Method and experiment}

\subsection{Hypothesis of particle identification using wavelength}
In this measurement, we identify the types of radiation particles detected based on the difference in scintillation wavelength between the emission of the host crystal and that of the dopant. Here we explain the luminescence mechanism of inorganic scintillators and the method of particle identification using them.

\begin{figure}[htbp]
 \begin{center}
 \includegraphics[width=13cm]{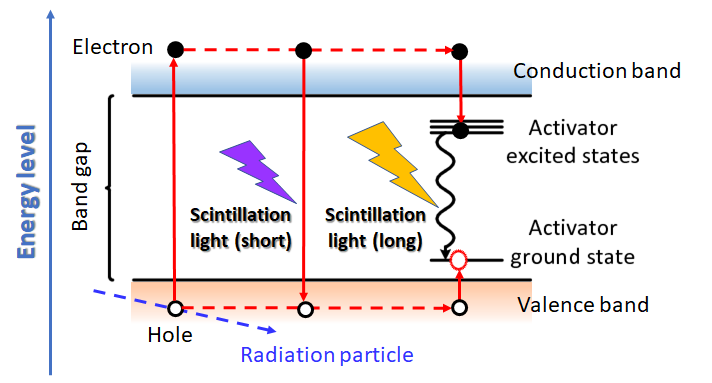}
\caption{Conceptual diagram of luminescence emission by inorganic scintillator host crystal and activator.}
\label{fig:scintillation}
 \end{center}
\end{figure}

Figure \ref{fig:scintillation} depicts the luminescence mechanism of inorganic scintillators. Inorganic scintillators have a crystal lattice structure, where the host crystal contains a few dopant atoms. Incident radiation particles excite electrons in the valence band, pushing them across the band gap into the conduction band, where electrons and holes are formed. When electrons and holes recombine, the energy corresponding to the band gap is emitted as scintillation light.
This is the scintillation light (short) emission in the figure. Generally, to efficiently recombine electrons and holes, activators are typically doped. This creates a new energy level inside the band gap, where the recombination of electrons and holes occur. In this case, light with an energy level corresponding to the energy level of the activator is emitted. Since this energy level is smaller than the band gap, light with lower energy and a longer wavelength is emitted as scintillation light.
In reality, the scintillator emission process is more complex and there are multiple emission processes, which will not be discussed in detail here (see \cite{1}).

Next, consider the case where the incoming radiation particles are electrons and alpha rays. Electrons have a small energy loss per unit distance (dE/dX), which results in a longer track length. As a result, since there are many activator atoms around the electron track, the main radiation will occur in the activator. In other words, the emitted wavelength will be long.
On the other hand, alpha rays have a shorter range because dE/dX is larger. Since more energy is deposited in a narrower area, the number of activator atoms relative to the excited electrons will be smaller. As a result, the luminescence at the activator is saturated, and more luminescence is expected to occur at the host crystal with alpha rays. In other words, the wavelength emitted is expected to be shorter in the alpha ray. Therefore, the emission wavelength may change depending on the type of particle, and this study aims to use this difference to identify particles.
The track lengths for beta and alpha rays were estimated using GEANT 4 simulation and it was about 1 mm for 1 MeV electrons and a several $\mu$m for 2 MeV $\alpha$-rays.

There are no measurements of the emission wavelengths of the Eu:LiCaI crystals used in this study. However, we know from our previous study that the undoped CaI$_2$ host emission peaks at 410 nm \cite{8}. Several host luminescence emissions have been observed for LiI crystals at 360, 414, and 458 nm \cite{9}. 
There are also literature values of 470 nm for Eu:CaI$_2$ and 475 nm for Eu:LiI \cite{10,11}. In other words, it is expected that the host and dopant luminescence can be separated at approximately 450 nm.

\subsection{Crystal preparation}
Because the difference in wavelength between the host and dopant luminescence is important in this measurement, the Eu concentration is expected to influence the measurement. Therefore, we prepared two types of Eu:LiCaI samples with different Eu concentrations.
Starting materials were prepared from powders of LiI enriched by $^6$Li (95 wt.\%), CaI$_2$, and EuI$_2$ (99.99\% purity, produced by APL). 
EuI$_2$:LiI:CaI$_2$ molar ratios of 1.5:25:73.5 for Sample A and B were used. 
Crystal growth was performed by the vertical Bridgman-Stockberger method in quartz ampoules with an inner diameter of 4 mm for sample A and 6 mm for sample B.

The weighed raw powder was put into a quartz ampoule in a glove box and heated under vacuum (1 Pa) at 300 $^{\circ}$C for at least 3 hours to remove residual oxygen and moisture. 
The ampoule was then sealed while being pulled into a high vacuum (10$^{-5}$ Pa). The ampoule was heated by a platinum heater and a high-frequency induction coil. The ampoule and platinum heater were surrounded by an alumina insulator to control the temperature gradient along the growth direction. The ampoule was pulled down at a rate of 0.2 mm/min. The grown crystals were mirror polished in a dry room with less than 2\% humidity to obtain circular samples 1 mm thick, and their luminescence and scintillation properties were measured. Details of the crystal growth setup are described in Ref \cite{8,12}.
The energy distribution of the Eu:LiCaI crystal was measured using a photomultiplier tube. A resolution of 17.2 \% (FWHM) at 662 keV was obtained.

No crystal structure data and phase diagram of the LiI-CaI$_2$ system have been reported yet. LiI and CaI$_2$ are strong deliquescence and are expected to have incongruent melt compositions, which would make single crystal growth difficult, and thus, the crystal information has not been reported. In this study, we report the neutron response performance of a new ternary eutectic composed of LiI, CaI$_2$, and/or these compounds.
The density of the crystals was not measured, but is expected to be about 4 g/cm$^3$ since the crystal densities of LiI and CaI$_2$ are 4.06 g/cm$^3$ and 3.96 g/cm$^3$, respectively.

\begin{figure}[htbp]
 \begin{center}
 \includegraphics[width=16cm]{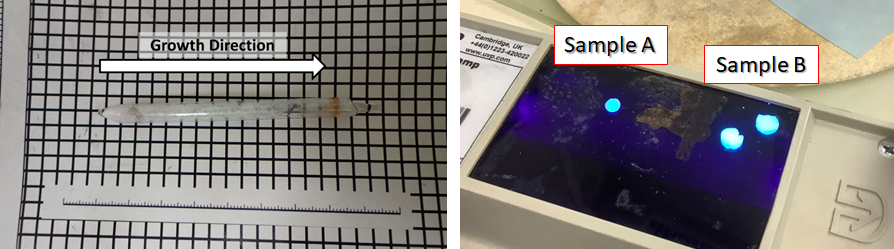}
\caption{(Left) The photograph of the as-grown Eu:LiCaI crystal (6mm $\phi$). The color of the crystal changes from white to green from the early to late stages of growth along the growth direction.(Right) Two samples of Eu:LiCaI crystal used in this measurement. The luminescence of Eu origin is visible by irradiating UV light from the bottom. 
}
\label{fig:crystal}
 \end{center}
\end{figure}

Figure \ref{fig:crystal} left shows the photograph of one of the grown crystals. The color of the crystal changes from white to green from the early to late stages of growth along the growth direction. It is due to the lower segregation coefficient of the Eu dopant. To compare samples with different Eu concentrations, sample A was cut from the white portion of the crystal and sample B was cut from the green portion.

Figure \ref{fig:crystal} right depicts a photograph of the two crystals used in this study irradiated with black light from the bottom; Sample A has a low Eu concentration and its luminescence is purple, and Sample B has a high Eu concentration and its luminescence is almost light blue. 
Compositional measurements by inductively coupled plasma mass spectrometry (ICP-MS, Agilent 8800, Agilent Technologies, CA, USA) yielded a difference of about two orders of magnitude in the concentration of Eu between Samples A (Eu 0.01\%) and B (Eu 4.74\%).

\begin{figure}[htbp]
 \begin{center}
 \includegraphics[width=12cm]{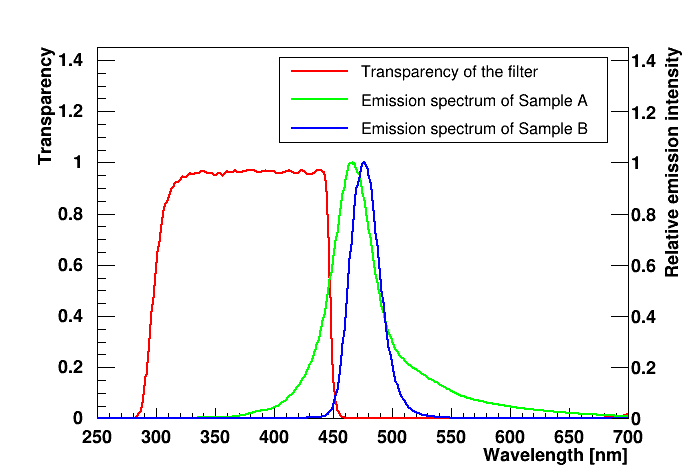}
\caption{Relative emission spectra of Sample A (green) and Sample B (blue), respectively. Transparency of the long cut filter used in this measurement is also shown (red).
}
\label{fig:Wavelength}
 \end{center}
\end{figure}

Figure \ref{fig:Wavelength} shows the measurement results of the emission wavelength spectrum of the actual sample by X-ray irradiation; the peak position is slightly higher and more sharply distributed in Sample B, which has a higher Eu concentration. In this sample, the light at 410 nm, which is the host emission of CaI$_2$, is almost nonexistent by X-rays. Sample A has tails on both sides of the longer and shorter wavelengths, showing a different emission spectrum from that of Sample B.

\subsection{Experimental setup and analysis}

\begin{figure}[htbp]
 \begin{center}
 \includegraphics[width=13cm]{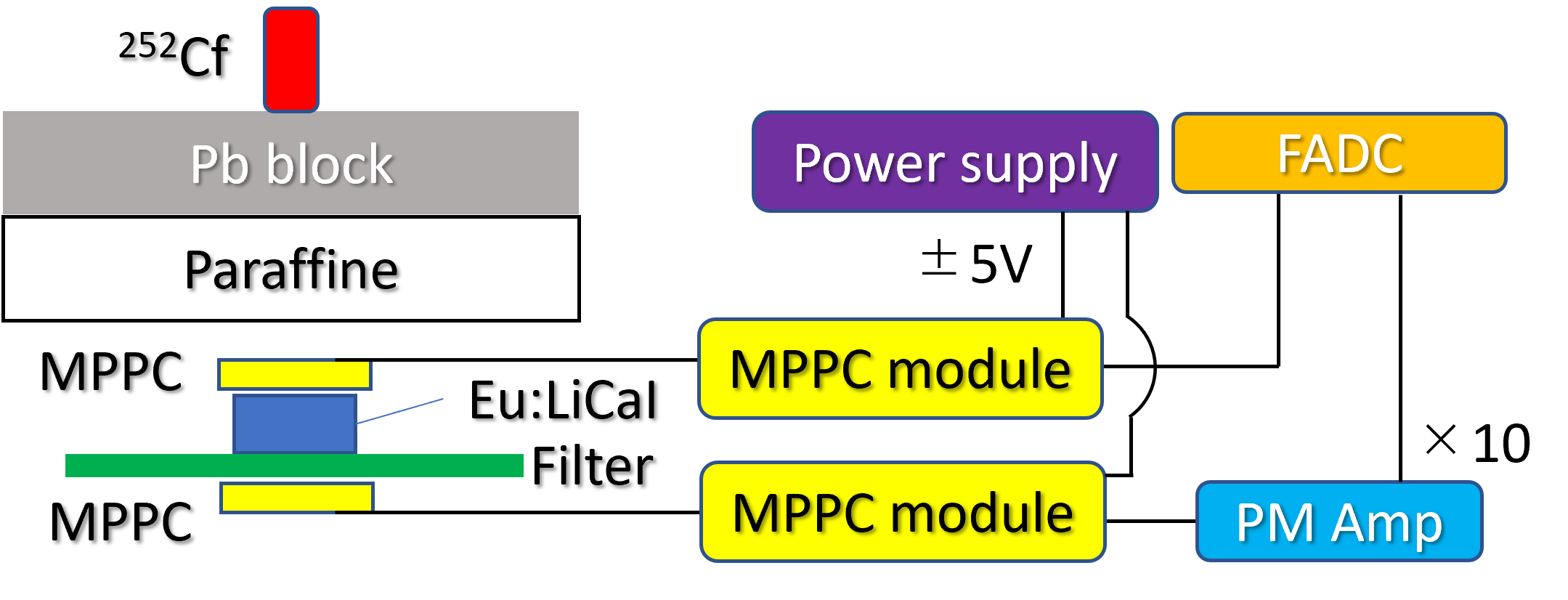}
\caption{Schematic of the experimental setup for this measurement. Size ratios differ from actual.
}
\label{fig:Setup}
 \end{center}
\end{figure}

Because Eu:LiCaI exhibits deliquescent properties, the experiment was conducted in a dry room maintained at about 2\% humidity.
Figure \ref{fig:Setup} depicts a schematic of the experimental apparatus with $^{252}$Cf source: Eu:CaI crystals were processed to about 1 mm thick and placed on a filter (Asahi Spectroscopy, SHX450) that cuts light at wavelengths above 450 nm. The transmittance of the filter is shown in Figure \ref{fig:Wavelength}.
Multi-Pixel Photon Counter (MPPC) module (Hamamatsu Photonics, C15524-3015SA) consisting of an MPPC, a preamplifier, a high-voltage power supply circuit, and a temperature compensation circuit were used for the measurement.
The output stability is $\pm$ 5\% in the catalog value, which is considered acceptable for this experiment.
The crystal and filter were sandwiched between two MPPCs, whose photosensitive surface was 3-mm square, at the end of a flexible cable.
An external low-voltage DC power supply (Matsusada, PLEE60-1.2) was used to drive the module by supplying ±5V.
The output signal from the MPPC is input to the Flash-Analog-to-Digital-Converter(FADC, CAEN, DT5720E). Because the signal from the MPPC on the filter side has a small output, it is amplified 10 times through a PM amplifier (REPIC, RPN-092) before being input to the FADC.
The sampling rate of the FADC was 250 MHz, and waveforms were acquired at 4 ns/ch. In addition, the trigger threshold of the FADC was set at 15 mV for unfiltered signals.
The sources used were $^{60}$Co (gamma source) and $^{252}$Cf (neutron source). $^{252}$Cf is a spontaneous fission source, emitting an average of 3.8 neutrons per decay and is used as a neutron source \cite{13}. To thermalize fast neutrons emitted from the source, paraffin was placed near the crystals.
Lead blocks were also placed to reduce gamma rays of spontaneous fission origin.
$^{60}$Co has placed approximately 5 cm away from the crystals, with nothing but a light-shielding sheet between them. Under these conditions, 10,000 event data were acquired for each source.

\begin{figure}[htbp]
 \begin{center}
 \includegraphics[width=16cm]{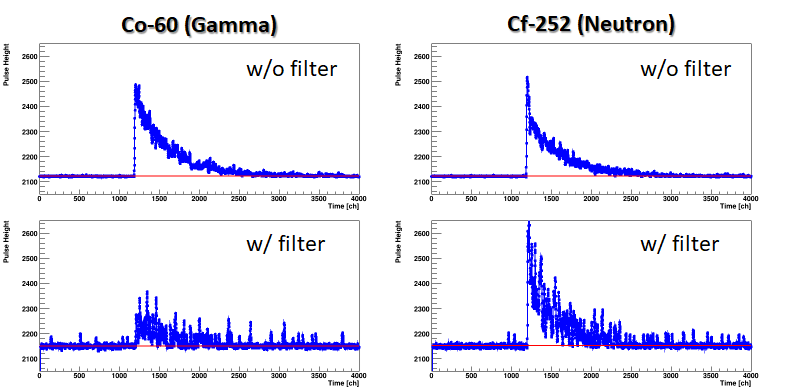}
\caption{Examples of waveforms obtained.
The event on the left is $^{60}$Co and the event on the right is $^{252}$Cf. The values on the vertical axis are plotted as they are for the FADC counts. Since the 2-volt range is read out in 12 bits, one count corresponds to almost 0.5 mV.}
\label{fig:Pulse}
 \end{center}
\end{figure}

Figure \ref{fig:Pulse} shows an example of a typical signal waveform for a $^{60}$Co, $^{252}$Cf event obtained with Sample B in this setup. The top waveforms are obtained without a filter, and the bottom waveforms are obtained with a filter. The unfiltered signal is thought to be originating from Eu-derived emission, whereas the filtered signal is thought to be originating from emission from the host crystal.
The two types of events, which are thought to originate from gamma and neutron radiation, detect almost the same amount of light without a filter, but there is a significant difference in their signal intensity when filtered. This indicates that the two types of events have different emission wavelengths.
The horizontal red line in the figure represents the pedestal calculated using the first 1,000 channels. In each event, the Analog-Digital converter (ADC) value is calculated by integrating the pedestal subtracted waveform over the 2,000 channels after the rise.
Two ADC values from different MPPCs are referred to as "filtered ADC" and "unfiltered ADC" with and without filter, respectively.

\section{Results}

\begin{figure}[htbp]
 \begin{center}
 \includegraphics[width=16cm]{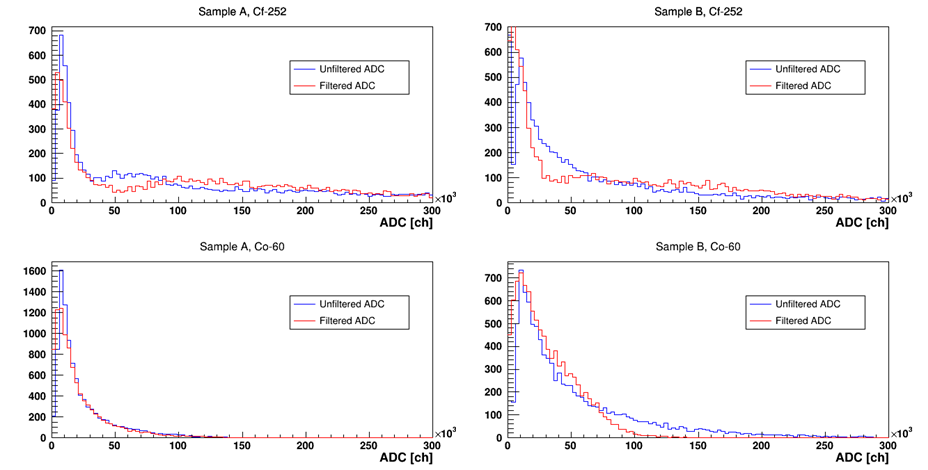}
\caption{One-dimensional distribution of ADC. Sample A is on the left, and Sample B is on the right, with $^{252}$Cf at the top and $^{60}$Co at the bottom. Red and blue indicate filtered and unfiltered ADCs, respectively.
}
\label{fig:ADC1d}
 \end{center}
\end{figure}

In this chapter, we investigate data using $^{60}$Co and $^{252}$Cf sources and discuss the possibility of particle discrimination by wavelength. Figure \ref{fig:ADC1d} depicts ADC one-dimensional distributions under different conditions. On the left is Sample A and on the right is Sample B. The upper panel is $^{252}$Cf and the lower panel is $^{60}$Co. The blue line is the unfiltered ADC and the red line is the filtered ADC.

For $^{60}$Co gamma rays, the distribution of filtered and unfiltered ADC is almost the same. This is presumably because the filter cuts out light at $\lambda>$450 nm and the PM amplifier circuit amplifies the signal by a factor of 10, resulting in almost identical distributions by chance.
The shape of the distribution is almost the same for the filtered/unfiltered ADC. The distribution for $^{252}$Cf shows the presence of a gamma-ray-like distribution at the low energy side. The distribution at the high energy side is also present, which seems to be $\alpha$+$^3$H event from the neutron capture reaction of $^{6}$Li.
The total energy of $\alpha$ and $^3$H is 4.78 MeV, which is a single energy, but because the size of the photosensor is smaller than the size of the crystal, the peak is greatly broadened due to the position dependence of the crystal.
Considering the distribution of Sample A, the filtered distribution shows a higher peak shift than the unfiltered distribution. This is probably because the emission wavelength of $\alpha$+$^3$H is shorter due to the larger emission ratio in the host crystal and light is less easily removed by the filter, and the signal is amplified by the PM amplifier, resulting in a shift toward higher peaks.
In Sample B, the unfiltered distribution appears slightly different, but this is probably because the peak of the $\alpha$+$^3$H distribution is buried by gamma rays because of the higher Eu concentration in Sample B and the higher light emission by gamma-ray.
In the filtered distribution, the peak of $\alpha$+$^3$H is shifted to the higher side and separated from the gamma-ray distribution.

Next, we define the ADC ratio as follows.
\begin{equation}
    ADC ratio = \frac{filtered \ ADC}{unfiltered \ ADC}
\end{equation}

\begin{figure}[htbp]
 \begin{center}
 \includegraphics[width=15cm]{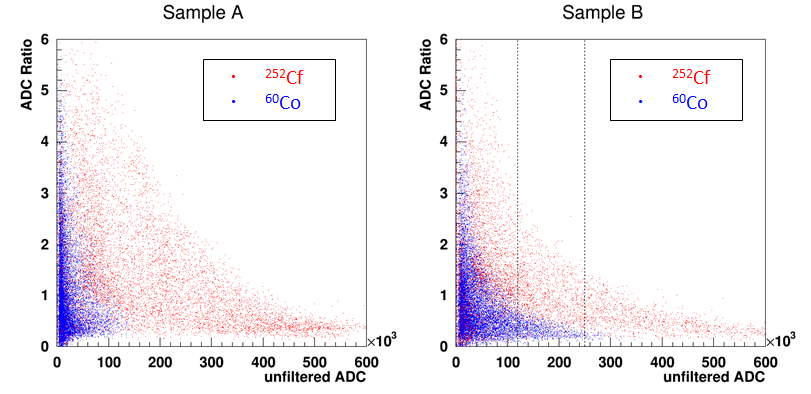}
\caption{Two-dimensional scatter plots of ADC ratio and unfiltered ADC for Sample A (left) and Sample B (right). Red and blue represent the data of $^{252}$Cf and $^{60}$Co, respectively.
}
\label{fig:ADCRatio2d}
 \end{center}
\end{figure}

Figure \ref{fig:ADCRatio2d} shows the ADC ratio as a function of unfiltered ADC. Red represents $^{252}$Cf, and blue represents $^{60}$Co.

There is a significant energy dependence, and the ratio tends to be smaller at higher energies for both neutrons and gamma rays.
However, when compared over the same energy range, the red $^{252}$Cf event always has a larger ADC ratio value than the blue $^{60}$Co event.
This indicates that the emission wavelength differs depending on the type of particle detected by Eu:LiCaI and that gamma rays and neutrons can be discriminated by the ADC ratio with and without a filter.
Comparing Samples A and B, the energy dependence of the ADC ratio is larger for Sample B with higher Eu doping, and the ADC ratio decreases rapidly for higher energy events.
Furthermore, Sample B has a larger light emission for $^{60}$Co gamma rays.

\begin{figure}[htbp]
 \begin{center}
 \includegraphics[width=10cm]{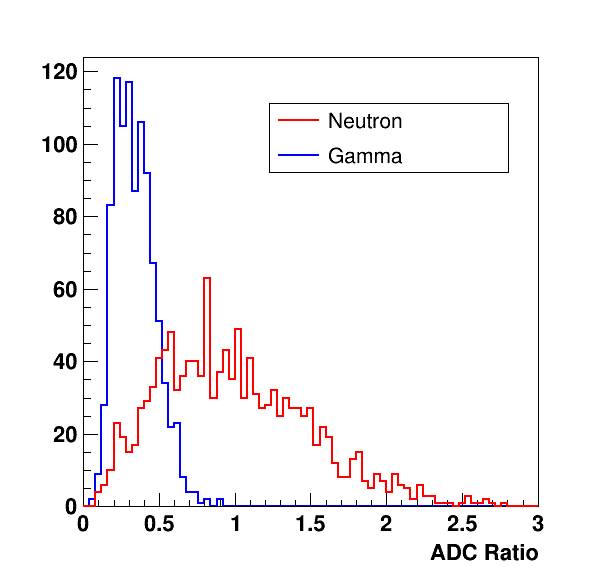}
\caption{One-dimensional distribution of ADC ratio for events in the range enclosed by the dotted line in Figure \ref{fig:ADCRatio2d}.
}
\label{fig:ADCRatio1d}
 \end{center}
\end{figure}

To visualize the separation performance between neutrons and gamma rays in the same energy region, the one-dimensional distribution of the ADC ratio for the events in the region indicated by the dotted line in Figure \ref{fig:ADCRatio2d} is shown in Figure \ref{fig:ADCRatio1d} for Sample B, where the light emission for gamma rays is large and easier to compare.
Most of the gamma rays illustrated in blue are distributed at ADC ratio $<$ 0.5, whereas the neutrons in red are distributed at higher ratios. In this energy region, it is possible to select neutron events with reduced background by excluding particles at ADC ratios $>$ 0.5.

\section{Discussion}
In the previous chapter, we showed the possibility to identify particles using wavelength information. In this chapter, we discuss whether the obtained results are correct and whether there is room for improving the separation performance.

\begin{figure}[htbp]
 \begin{center}
 \includegraphics[width=16cm]{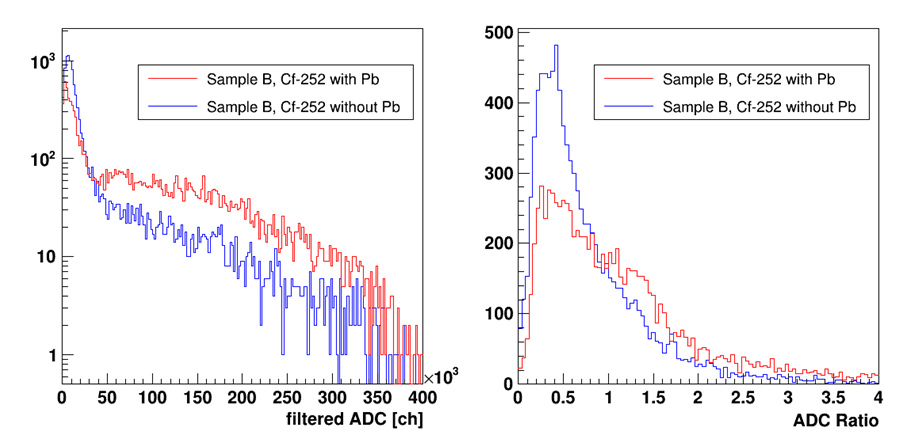}
\caption{ADC distribution with/without lead (left) and ADC ratio with/without lead (right) when irradiated with $^{252}$Cf source.
}
\label{fig:PbOnOff}
 \end{center}
\end{figure}

In the Results section, we demonstrated that the data obtained with the $^{252}$Cf source exhibited a gamma-ray-like distribution at lower energies and $\alpha$+$^3$H-like events at higher energies. To confirm whether this is true, we compared data acquired with and without a lead block placed between the $^{252}$Cf source and the crystal.
The left panel of Figure \ref{fig:PbOnOff} depicts the filtered ADC distribution with lead in red and without lead in blue.
The without-lead distribution is dominated by low-energy events, which are reduced by the presence of Pb. Because these events are attenuated by lead, the low-energy events are probably gamma rays originating from $^{252}$Cf.
The increase in events on the high energy side due to the placement of lead is thought to be a relative increase caused by the decrease in gamma rays because neutrons are not shielded by lead.

Next, the one-dimensional distribution of the ADC ratio with and without lead is compared in Figure \ref{fig:PbOnOff} right. All events are included without energy selection.
In the red distribution with lead, only the area with a small ADC ratio has decreased, whereas the area with a large ADC ratio has increased relatively more than the blue distribution with no lead.
The events with a small ADC ratio that decreased with the placement of lead are considered gamma rays from $^{252}$Cf. This indicates that the ADC ratio can be used to identify particles.

\begin{figure}[htbp]
 \begin{center}
 \includegraphics[width=10cm]{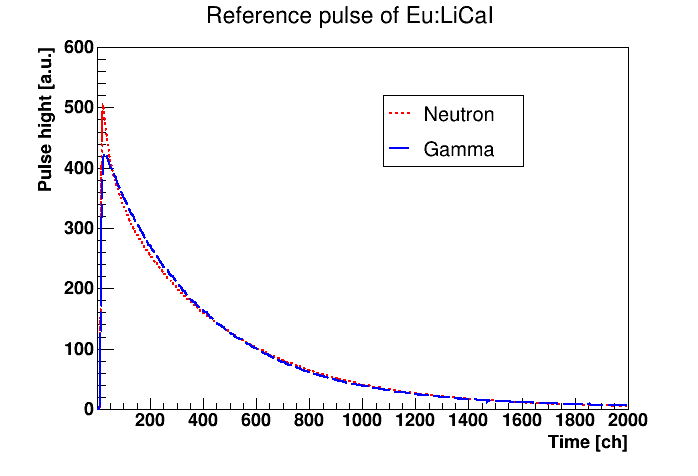}
\caption{Average waveforms for neutron and gamma events.
}
\label{fig:refpulse}
 \end{center}
\end{figure}

Next, we investigated the differences in the response waveforms of each particle. If there are differences, it would further improve the performance of identifying particles.
Figure \ref{fig:refpulse} shows the averaged waveforms of selected neutron and gamma events from the data.
The neutron events in red include the events with high filtered ADC in the $^{252}$Cf data.
For the gamma-ray events in blue, the $^{60}$Co data was used with no cuts applied.
There is no significant difference in the time constant of the waveform between neutrons and gamma rays. 
However, the beginning of the waveform is slightly sharper for events of neutron origin. 
If this slight difference can be incorporated into the analysis, there is room for improving particle discrimination.

In this study, we investigated whether the wavelength difference between the host and activator luminescence could be used for particle identification. Differences were obtained for different sources of radiation, and the results support our hypothesis.
However, another possibility is that the Eu:LiI and Eu:CaI$_2$ phases exist separately, and the difference may have manifested itself.
In other words, short-range alpha rays emitted from $^6$Li emit light only at Eu:LiI.
Gamma rays also emit light in Eu:CaI$_2$ because of their long range, raising the possibility that we are detecting the difference in wavelength between the two types of crystals.
We will attempt to fabricate a single LiCa$_2$I$_5$ crystal and perform similar measurements in the future. We will also perform detailed crystal structure analysis by XRD measurements and transmission electron microscopy.

Incidentally, the same experiment and analysis have also been performed on Eu:LiCAF crystals \cite{2}, which also contain $^6$Li. The results showed no difference between neutrons and gamma rays. This indicates that our analytical method is not biased and that the wavelength difference is visible in Eu:LiCaI.

\section{Conclusion}

In this study, we investigated a particle identification technique using scintillation wavelength information in Eu:LiCaI crystals. This was based on the fact that the ratio of host to activator emission differed depending on the type of incident particle due to the difference in energy loss per unit distance.
The experimental results showed that the ADC distribution was changed when irradiated with $^{252}$Cf by passing a filter of $\lambda >$ 450 nm. On the other hand, no significant difference in the ADC distribution was observed when irradiating with a $^{60}$Co source, and the ADC ratio with and without a filter differed depending on the source. We found for the first time that ADC ratios could be used to identify the type of particles detected.

These findings also indicate that the Eu:LiCaI crystal is a promising future neutron detector containing $^6$Li. In addition, our wavelength-based particle identification method is highly versatile and can be applied to other crystals. In the future, further detailed verification of the results, investigation of crystals that can be used for wavelength discrimination, optimization of Eu concentration and Li/Ca ratio, etc., should be conducted.

\section*{Acknowledgment}

This work was supported by Shimadzu Science and Technology Foundation, the Asahi Glass Foundation.
This work was partially supported by JSPS KAKENHI Grant Numbers 18H01222 and 22H04570.

\end{document}